\documentclass[preprint2]{aastex}
\usepackage{url}

\newcommand{\ma}{2XMMi J003833.3+402133}
\newcommand{\cf} {XMM J0038+40}
\newcommand{\xmm}{\emph{XMM-Newton}}

\slugcomment{}

\shorttitle{A Candidate Magnetar at High Galactic Latitude}
\shortauthors{Callingham et al.}

\begin{document}


\title{The X-ray Transient 2XMMi J003833.3+402133: A Candidate Magnetar at High Galactic Latitude}


\author{J. R. Callingham\altaffilmark{1}, S. A. Farrell\altaffilmark{1,2}, B. M. Gaensler\altaffilmark{1,2}, G. F. Lewis\altaffilmark{1}, and M. J. Middleton\altaffilmark{3}}
\affil{$^1$Sydney Institute for Astronomy (SIfA), School of Physics, The University of Sydney, NSW 2006, Australia}
\affil{$^2$ARC Centre of Excellence for All-sky Astrophysics (CAASTRO)}
\affil{$^3$Department of Physics, University of Durham, South Road, Durham DH1 3LE, UK}
\email{j.callingham@physics.usyd.edu.au}



\begin{abstract}
We present detailed analysis of the transient X-ray source \ma\ detected by \xmm\ in January 2008 during a survey of M 31. The X-ray spectrum is well fitted by either a steep power law plus a blackbody model or a double blackbody model. Prior observations with \xmm, \emph{Chandra}, \emph{Swift} and \emph{ROSAT} spanning 1991 to 2007, as well as an additional \emph{Swift} observation in 2011, all failed to detect this source. No counterpart was detected in deep optical imaging with the Canada France Hawaii Telescope down to a 3$\sigma$ lower limit of $g$ = 26.5 mag. This source has previously been identified as a black hole X-ray binary in M 31. While this remains a possibility, the transient behaviour, X-ray spectrum, and lack of an optical counterpart are equally consistent with a magnetar interpretation for 2XMMi J003833.3+402133. The derived luminosity and blackbody emitting radius at the distance of M 31 argue against an extragalactic location, implying that if it is indeed a magnetar it is located within the Milky Way but 22$\degr$ out of the plane. The high Galactic latitude could be explained if \ma\ were an old magnetar, or if its progenitor was a runaway star that traveled away from the plane prior to going supernova.
\end{abstract}


\keywords{Stars: Neutron, X-rays: individual (2XMMi J003833.3+402133), X-rays: stars, Galaxies: individual: M 31}



\section{Introduction}
Magnetars are a rare class of neutron stars which are characterised by a very strong surface magnetic field, typically 10$^{14}$ -- 10$^{15}$ G \citep{dun92,woo06}, and relatively slow spin rate (P $\sim$2 -- 12 s) when compared to the majority of observed pulsars. They also differ from normal pulsars in that they are not powered by their spin-down energy losses but by the decay of energy stored in their magnetic field \citep{dun92}. 


Magnetars are historically divided into two classes: anomalous X-ray pulsars (AXPs) which are characterised by slow pulsations and rapid spin-down, and soft gamma repeaters (SGRs) which exhibit sporadic bursts of hard X-ray/soft gamma rays. While originally considered to be unrelated, it is now accepted that AXPs and SGRs are the same type of object \citep{kou98,gav02,kas03}. There are currently 16 confirmed magnetars (7 SGRs and 9 AXPs) and a further 7 candidates (4 SGRs and 3 AXPs)\footnote{From McGill's Online Catalogue available at \url{http://www.physics.mcgill.ca/~pulsar/magnetar/main.html}.}. Due to this small sample size it is difficult to determine the average lifetime, spatial distribution, duty cycle and luminosity of the population as a whole. However, all the Galactic magnetars lie very close to the Galactic plane, consistent with young ($<$10,000 yrs) neutron stars with low spatial velocities \citep[$<$500km s$^{-1}$;][]{gae01}. The exceptions are the two magnetars located in the Magellanic Clouds \citep{maz79,mcg05}. While there have been reports of the detection of SGRs in M 31 \citep{ofe08,maz08} and M81 \citep{fre07}, these sources have been detected only once and their SGR nature has thus yet to be verified.

In this paper we present an in-depth analysis of multi-wavelength data on the X-ray source \ma\ (hereafter, referred to as \cf), which was observed during a survey of the nearby galaxy M 31. \citet{sti11} identified this object (source 57 in their sample) as a probable black hole low-mass X-ray binary (LMXB) located within M 31. Here we argue that the X-ray spectral and timing properties of this source are also consistent with \cf\ being a new addition to the rare class of Galactic magnetars. If this conclusion is correct, the high Galactic latitude of \cf\ would make it unique as the other Galactic magnetars are all located within the plane. 


\section{Data Reduction, Analysis \& Results}
\subsection{X-ray Data}

The field of \cf\ was observed by \xmm\ on 2008 January 2 for 45.5 ks by the three EPIC cameras (pn, MOS1 and MOS2) as part of a survey of the local group galaxy M 31 \citep{sti11}. Source detection performed by the \xmm\ processing pipeline detected a bright uncatalogued source at an off-axis angle of 7.72$\arcmin$. The source position is RA = 00h38\arcmin 33.32\arcsec, dec = +40\degr21\arcmin33.20$\arcsec$ (J2000) with a 3$\sigma$ positional uncertainty radius of 1$\arcsec$ (see \citet{wat09} for a detailed description of the source detection procedure). No data were available from the two Reflection Grating Spectrometers (RGS) as \cf\ fell outside their fields of view. 

\begin{figure}
\begin{center}
\includegraphics[angle=-90,width=\columnwidth]{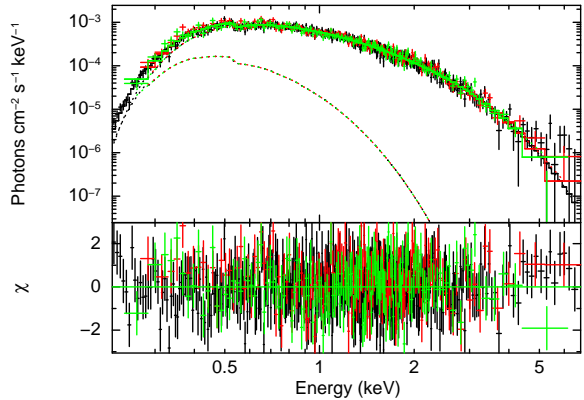}
 \caption{\xmm\ unfolded EPIC spectra (black = pn, red = MOS1, green = MOS2) of \cf\ from the 2008 detection fitted with the Phabs*(DISKBB+CompTT) model. The bottom panel shows the fit residuals.}\label{dbbplmod}
\end{center}
\end{figure}

\begin{figure}
\begin{center}
\includegraphics[angle=-90,width=\columnwidth]{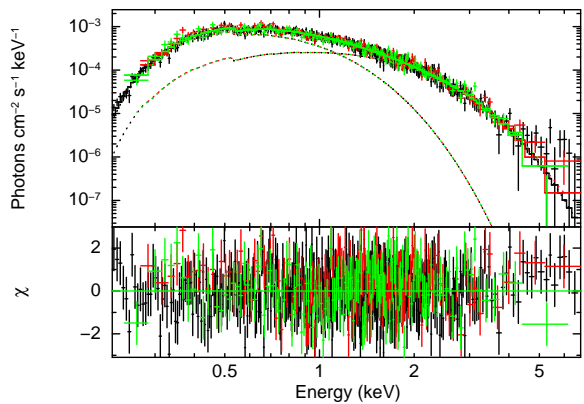}
 \caption{\xmm\ unfolded EPIC spectra (black = pn, red = MOS1, green = MOS2) of \cf\ from the 2008 detection fitted with the Phabs*(BB+BB) model. The bottom panel shows the fit residuals.}\label{specbb}
\end{center}
\end{figure}

The Science Analysis System (SAS) version 10.0 software\footnote{\url{http://xmm.esa.int/sas/}} was used to process the Observation Data Files (ODFs) using the same procedure as outlined by \citet{far10}. Background light curves extracted from the entire field at energies $>$ 10 keV found no evidence for flaring, and so no good time interval filtering was applied. We extracted source spectra using circular regions centred on the source position with radii of 71.5\arcsec\ and 41.5\arcsec\ for the pn and MOS cameras, respectively\footnote{The source extraction radii were chosen so as to maximise the signal-to-noise ratio of the data products (V. Braito 2011, private communication).}. Background spectra were extracted from nearby source-free regions of the field with areas 1.5 times the source extraction region areas. Response and ancillary response files were produced for the spectra and the light curves were corrected using the SAS task \emph{epiclccorr}. The spectra were grouped at 20 counts per bin to provide sufficient statistics for $\chi^2$ fitting and analysed using XSPEC v12.6.0q \citep{arn96}. Data below 0.2 keV and above 7 keV (where the statistics are very poor) were ignored for the spectral fitting. 

\subsubsection{X-ray Spectral Analysis}
\citet{sti11} fitted spectra extracted from their 2008 \xmm\ observation with three models: absorbed disk blackbody plus power law, absorbed disk blackbody, and absorbed bremsstrahlung models. They found that the disk blackbody plus power law model was the best fitting model obtaining $\chi^2$/dof = 174/145, although they also claim formally acceptable fits using the two other models (with $\chi^2$/dof = 270/147 and 209/147 for the simple disk blackbody and bremsstrahlung models, respectively). 

We fitted the spectra we extracted from the 2008 \xmm\ observation with the same models and obtained similar results for the disk blackbody plus power law and the simple disk blackbody models, although we argue that due to the high quality of the data the fit with the simple disk blackbody model is unacceptable with a $\chi^2$/dof = 812/714. We therefore only report the results of the disk blackbody plus power law fit in Table \ref{bhxrb}. In contrast, our fit with the absorbed bremsstrahlung model obtained a much better fit ($\chi^2$/dof = 764/714, see Table \ref{specpar}) with a significantly lower temperature of kT = 0.98 $\pm$ 0.02 keV than the kT = 1.91 $\pm$ 0.07 keV obtained by \citet{sti11}. The cause of the differences between our fitting and that reported by \citet{sti11} is unclear. We also fitted the spectra with a more physical model representing emission from an accreting black hole, i.e. an absorbed disk blackbody plus thermal Comptonisation model (compTT in XSPEC), with the input soft photon (Wien) temperature fixed to the disk blackbody temperature. Again, we obtained an acceptable fit with $\chi^2$/dof = 746/711 (see Table \ref{bhxrb}). 



In addition to the model fits reported by \citet{sti11}, we attempted to fit the 2008 \xmm\ spectra with power law, blackbody (the BBODYRAD model in XSPEC), and thermal plasma \citep[the MEKAL model in XSPEC;][]{mew85,mew86,kaa92,lie95} models. In each case photoelectric absorption was accounted for using the phabs component in XSPEC and the Wilms abundances \citep{wil00}. Neither the simple power law model nor blackbody models provided an acceptable fit, with $\chi^2$/dof = 1205/719 and 1165/719, respectively, and significant residuals appearing below 2 keV. Adding a low temperature blackbody component to the power law model improved the fit significantly ($\chi^2$/dof = 806/712), although with a very steep power law photon index (see Table \ref{specpar}). The addition of a second blackbody component to the simple absorbed blackbody model produced a better fit ($\chi^2$/dof = 752/712) and completely smoothed out the low energy residuals (see Table \ref{specpar}).  

Attempts to fit the spectra with an absorbed MEKAL model with the abundance parameter frozen at Solar values did not provide an acceptable fit ($\chi^2$/dof = 11527/719). Allowing the abundance to vary freely improved the fit significantly ($\chi^2$/dof = 769/718), however the abundance value fell to zero indicating that no significant line emission is present and therefore the model is consistent with the underlying bremsstrahlung continuum model. In summary, we obtained acceptable fits with the double blackbody, blackbody plus power law, disk blackbody plus power law, disk blackbody plus thermal Comptonisation, and bremsstrahlung models but are able to rule out the simple blackbody, simple power law, and thermal plasma (with non-zero elemental abundances) models. The best fits were obtained with the disk blackbody plus thermal Comptonisation and the double blackbody models, which are shown in Figures \ref{dbbplmod} and \ref{specbb}.

\subsubsection{X-ray Timing Analysis}
To search for pulsations in the \xmm\ data, we extracted counts in the energy range 0.2 -- 12~keV using the same extraction regions as used for the spectra. This resulted in 8078, 7192 and 23740 counts for the MOS1, MOS2 and pn cameras, respectively. In each case, we corrected the photon arrival times to the Solar system barycenter. To search for high frequency periodic variability we performed a $Z^2_1$ (Rayleigh) test to search for sinusoidal pulsations \citep[e.g.][]{lea83}. For the MOS1 and MOS2 data, we searched for pulse periods in the range 6.4 to 4500~s, while for pn data we considered periods between 146~ms and 4500~s. We saw no significant power at any period, either in each data set separately or in a period search of all three data sets combined. The corresponding 5$\sigma$ upper limit on the pulsed fraction is 7\% in the period range 146~ms to 6.4~s, and 5.5\% in the period range 6.4 to 4500~s.

We also extracted power density spectra (PDS) over the reliable energy bandpass (0.3-6~keV) binned on 10 s. We find that, with linear binning, the variability is consistent with the statistical white noise within 1$\sigma$ using the most accurate Bayesian least-squares fitting routines \citep{vau10}. As the shape of the PDS is energy dependent and we could conceivably have a stable, bright component diluting the variability when averaged across the entire bandpass, we also extracted the energy dependent fractional excess variance \citep[see e.g.][]{ede02,vau03}. We did this initially for the bands which correspond to the absorbed disk blackbody plus thermal Comptonisation spectral model (0.3 -- 0.8 and 0.8 -- 6keV; see Figure \ref{dbbplmod}) with binning on 25, 50, 100, 200 and 400 s to test for variability typically seen from black hole X-ray binaries. We could not constrain any variability over any timescales in either energy band above the white noise level, obtaining a 3$\sigma$ upper limit of 8$\%$ on the fractional variability\footnote{We note however that this does not take into account the effect of red noise.} using the method of \citet{van89}. This is the theoretical upper limit for what we should be able to detect given teh bandpass, count rate, and exposure time. As this relies on our spectral deconvolution being an accurate description of the data we also tested energy bands within these two bands (specifically 0.3 -- 0.5, 0.5 -- 1, 1 -- 2 and 2 -- 6 keV) on 400 s binning (to ensure adequate statistics in these smaller energy bands) and find no constrained variance above the statistical white noise. 

\begin{table*}
\begin{center}
\caption{Best fit spectral parameters to the 2008 \xmm\ observation of \cf\ fitted with disk blackbody plus power law, and disk blackbody plus thermal Comptonisation models.\label{bhxrb}}
\begin{tabular}{cccc}
\tableline\tableline
Parameter & DISKBB+Pow\tablenotemark{a} & DISKBB+CompTT\tablenotemark{a}  & Units\\
\tableline
N$_H$ & 0.13$^{+0.1}_{-0.02}$ & 0.11$^{+0.05}_{-0.03}$ & 10$^{22}$ cm$^{-2}$\\
$kT_1$\tablenotemark{b} & 0.46$\pm$0.01 & 0.17$^{+0.2}_{-0.02}$ & keV\\
Norm$_1$\tablenotemark{c} & 2.2$^{+0.2}_{-0.3}$ & 16$^{+39}_{-16}$& \nodata \\
$\Gamma$/$T_0$\tablenotemark{d} & 2.4$^{+0.9}_{-0.5}$ & 0.17$^{+0.2}_{-0.02}$ & \nodata/keV \\
$kT_2$\tablenotemark{e} & \nodata & 0.50$\pm$0.03 & keV\\
$\tau$\tablenotemark{f} & \nodata &14$^{+2}_{-1}$& \nodata \\
Norm$_2$\tablenotemark{g} & 8$^{+30}_{-5} \times 10^{-5}$ & 4$^{+1}_{-3} \times 10^{-3}$  & \nodata\\
Flux$_{abs}$\tablenotemark{h} & 1.53$\pm$0.02 & 1.48$\pm$0.02 & 10$^{-12}$ erg cm$^{-2}$ s$^{-1}$\\
Flux$_{unabs}$\tablenotemark{h} & 2.7$^{+3.0}_{-0.4}$ & 2.08$^{+0.5}_{-0.3}$ &  $10^{-12}$ erg cm$^{-2}$ s$^{-1}$ \\
Flux$_1$/Flux$_2$\tablenotemark{i} & 2.37 &   0.10&  \nodata\\
$\chi^2$/dof & 778/712 & 746/711  & \nodata \\
\tableline
\end{tabular}
\tablenotetext{a}{All models include an absorption component. Errors are quoted at the 90$\%$ confidence level.}
\tablenotetext{b}{Temperature of the first model component.}
\tablenotetext{c}{Normalisation of the first model component.}
\tablenotetext{d}{This row contains the photon index for the power law component in the DISKBB+POW model and the soft photon temperature ($T_0$) for the thermal Comptonisation component in the DISKBB+CompTT model.}
\tablenotetext{e}{Plasma temperature of the thermal Comptonisation model component.}
\tablenotetext{f}{Plasma optical depth.}
\tablenotetext{g}{Normalisation of the second model component.}
\tablenotetext{h}{Absorbed and unabsorbed fluxes are calculated over the energy range 0.2 -- 10 keV.}
\tablenotetext{i}{Ratio of unabsorbed 0.2 -- 10 keV fluxes of model component 1 over model component 2.}
\end{center}
\end{table*}

\begin{table*}
\begin{center}
\caption{Best fit spectral parameters to the 2008 \xmm\ observation of \cf\ fitted with the bremsstrahlung, double blackbody and blackbody plus power law models.\label{specpar}}
\begin{tabular}{ccccc}
\tableline\tableline
Parameter &Bremsstrahlung & BB+BB\tablenotemark{a} & BB+Pow\tablenotemark{a} & Units\\
\tableline
N$_H$ &  0.194$\pm$0.006 & 0.08$\pm$0.01 & 0.35$^{+0.02}_{-0.01}$ &   10$^{22}$ cm$^{-2}$\\
$kT_1$\tablenotemark{b} & 0.98$\pm$0.02 & 0.207$\pm$0.005 & 0.37$\pm$0.01 &  keV\\
Norm$_1$\tablenotemark{c} & 1.72$\pm$0.06 $\times$ 10$^{-3}$ & 48$^{+18}_{-11}$ & 4.1$^{+1.0}_{-0.7}$  & \nodata \\
$kT_2$/$\Gamma$\tablenotemark{d} & \nodata& 0.44$^{+0.01}_{-0.02}$ & 3.7$^{+0.2}_{-0.1}$  & keV/\nodata \\
Norm$_2$\tablenotemark{e} & \nodata& 2.5$^{+0.7}_{-0.6}$ & 7.8$^{+0.7}_{-0.8} \times 10^{-4}$  &  \nodata\\
Flux$_{abs}$\tablenotemark{f} & 1.51$\pm$0.02 & 1.50$\pm$0.02 & 1.50$\pm$0.02 &  10$^{-12}$ erg cm$^{-2}$ s$^{-1}$\\
Flux$_{unabs}$\tablenotemark{f} & 3.18$^{+0.07}_{-0.06}$& 1.90$^{+0.10}_{-0.08}$ & 12$^{+3}_{-2}$ &  $10^{-12}$ erg cm$^{-2}$ s$^{-1}$ \\
Flux$_1$/Flux$_2$\tablenotemark{g} & \nodata& 0.94 & 0.07  &  \nodata\\
$\chi^2$/dof & 764/714 & 752/712 & 806/712 & \nodata \\
\tableline
\end{tabular}
\tablenotetext{a}{All models include an absorption component. Errors are quoted at the 90$\%$ confidence level.}
\tablenotetext{b}{Temperature of the first model component.}
\tablenotetext{c}{Normalisation of the first model component.}
\tablenotetext{d}{This row contains the temperature for the second BB component in the BB+BB model and the photon index of the power law component in the BB+Pow model.}
\tablenotetext{e}{Normalisation of the second model component.}

\tablenotetext{f}{Absorbed and unabsorbed fluxes are calculated over the energy range 0.2 -- 10 keV.}
\tablenotetext{g}{Ratio of unabsorbed 0.2 -- 10 keV fluxes of model component 1 over model component 2.}
\end{center}
\end{table*}


\subsubsection{Additional X-ray Observations}
M 31 has been observed numerous times previously in X-rays, and thus archival data from \emph{ROSAT}, \xmm, \emph{Chandra}, and \emph{Swift} were also analysed. In addition, following the discovery of \cf\ we obtained Target of Opportunity (ToO) observations of the field with the \emph{Swift} X-ray Telescope (XRT) between 04 -- 15 February 2011. For the additional \xmm\ data, we extracted images from all three EPIC cameras after filtering out background flares. Pre-processed images from the archival \emph{Chandra}, \emph{ROSAT}, and \emph{Swift} data were downloaded from NASA's HEASARC archive\footnote{\url{http://heasarc.gsfc.nasa.gov/cgi-bin/W3Browse/w3browse.pl}}. The 2011 \emph{Swift} ToO observation was processed using the online XRT data processing facility\footnote{\url{http://www.swift.ac.uk/user_objects/}} \citep{eva09}. 

\begin{figure*}
\epsscale{2.1}
\plotone{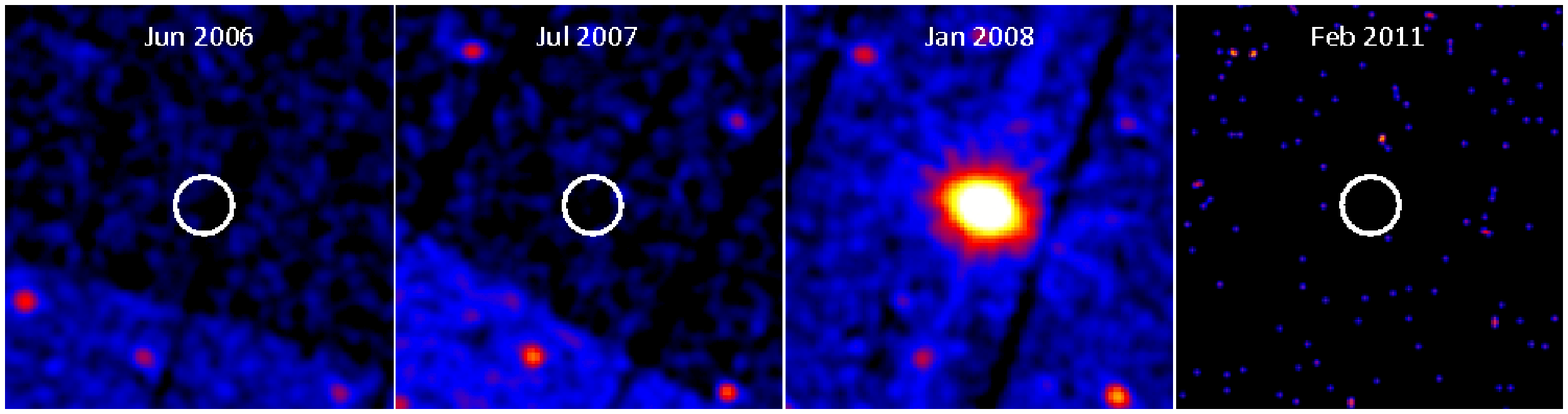}
\caption{X-ray images of the field of \cf\ taken with \xmm\ (panels 1 to 3 from the left) and \emph{Swift}. The white circle indicates the position of \cf\ with a radius of 30\arcsec. The \xmm\ images span the energy range of 0.2 -- 12 keV, while the {\em Swift} image spans 0.3 -- 10 keV. A Gaussian smoothing function with a kernel radius of 3 pixels was applied to all images. \label{ximages}}
\end{figure*}

 \cf\ was not detected in any of these data (Figure \ref{ximages}). Upper limits were estimated using the count rates measured within regions centred on the position of \cf\ in each of the images, subtracted from background count rates measured in nearby source-free regions. The WebPIMMS online flux conversion tool\footnote{\url{http://heasarc.gsfc.nasa.gov/Tools/w3pimms.html}} and the double blackbody spectral model obtained from fitting the EPIC spectra from the 2008 \xmm\ observation were used to convert the 3$\sigma$ count rate upper limits into flux limits. The flux limits derived in this manner are all well below the flux observed in the 2008 \xmm\ observation indicating that \cf\ is a transient source with variability by a factor $\gtrsim$ 450. Specific details of the observations from which data was analysed including the upper limits of the non-detections are given in the online Appendix (Table \ref{obslog}).

\subsection{UV, Optical and Near-Infrared Data}
\cf\ fell within the field of view of the Optical Monitor (OM) during all three \xmm\ observations, which observed the field in imaging mode with the \emph{uvw1} and \emph{uvm2} UV filters. As the 2008 OM images are the only available UV data that were taken at the time when \cf\ was X-ray bright, we focused on these data for our analysis (though we note that no source was detected within the X-ray error circle in any of the other OM or \emph{Swift} UVOT data). The total exposure times during this observation were 1461 s and 6317 s for the \emph{uvw1} and \emph{uvm2} filters, respectively. No source was detected within the 3$\sigma$ X-ray error circle of \cf\ in any of the images. We determined 3$\sigma$ count rate upper limits at the position of \cf\ using the \xmm\ pipeline processed images and the same method as that used for the X-ray non-detections, and converted the count rates into Vega magnitudes using the filter zero points for the OM\footnote{\url{http://xmm.esa.int/external/xmm_user_support/documentation/uhb_2.5/node75.html}}.
 
\begin{figure}
\begin{center}
\includegraphics[width=\columnwidth]{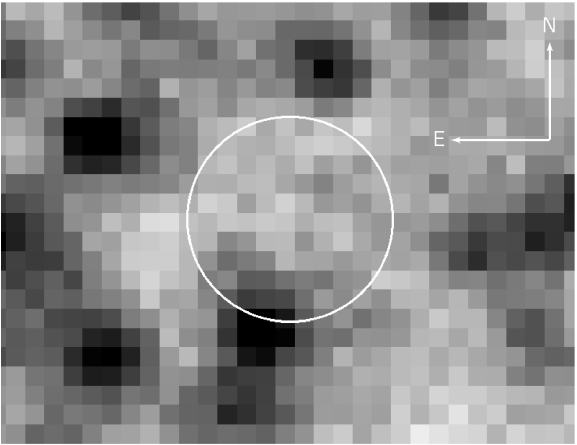}
 \caption{CFHT $g$-band image with the X-ray position of \cf\ indicated by the white circle, with a radius equal to the 3$\sigma$ positional error of 1\arcsec.}\label{gband}
\end{center}
\end{figure}

Deep $g$- and $i$-band images of the field of \cf\ were taken with the Canada France Hawaii Telescope (CFHT) on 2008 October 3 as part of the Pan-Andromeda Archaeological Survey \citep[PAndAS;][]{mcc09}. The CFHT observations and data reduction procedures are described by \citet{iba07}. No counterpart was detected in either band image within the X-ray error circle of \cf\ (see Figure \ref{gband}), so we determined magnitude lower limits in the same manner as described above. There is however a faint point source just outside and to the South-East of the error circle at RA = 00h38\arcmin33.3\arcsec, dec = +40$^\circ$21\arcmin31.9\arcsec\ with $g$ = 24.3 $\pm$ 0.1 mag and $i$ = 22.11 $\pm$ 0.04 mag which could possibly be the counterpart. No source was present within or even nearby the X-ray error circle in the 2MASS J, H, and K$_s$ images. In order to estimate conservative magnitude lower limits in the 2MASS near-infrared (NIR) data, we identified the faintest source in each band within 30\arcmin\ of \cf\ and adopted these magnitudes as the lower limits. All the lower limits derived from the UV, optical and NIR data are given in Table \ref{optlim}.

\begin{table}
\begin{center}
\caption{UV, Optical and NIR Lower Limits}\label{optlim}
\begin{tabular}{cccc}
\tableline\tableline
Telescope/ & Filter & Date & Lower Limit\\
Survey& & & (mag)\\
\tableline
OM & \emph{uvw1} &2008-01-02 & 20.1\\
OM &\emph{uvm2} & 2008-01-02 & 19.3\\
CFHT & \emph{g} &2008-10-03 & 26.5 \\
CFHT & \emph{i} &2008-10-03 & 25.5 \\
2MASS &  J & 2000-10-25 &17.7 \\
2MASS &  H & 2000-10-25 &16.6 \\
2MASS &  K$_s$ & 1998-12-04 &16.3\\
\tableline
\end{tabular}
\end{center}
\end{table}

\subsection{Radio Data}
The region surrounding \cf\ has been imaged at various radio frequencies as part of surveys of M 31. Using these data, we find no radio source at the position of \cf\ at frequencies of 0.3, 0.6 or 1.4~GHz, with 5$\sigma$ upper limits on the flux of a radio counterpart of 5, 4 and 1.5~mJy, respectively \citep{bys84,wal85,gel04}. Less sensitive observations at 0.3 and 1.4~GHz at other epochs also show no radio source at this position \citep{ren97,con98}.

\section{Discussion \& Conclusions}


The X-ray spectrum, transient behaviour, and lack of an optical and infrared counterpart place strong constraints on the nature of \cf. Non-transient objects such as isolated cooling neutron stars and rotation powered pulsars with moderate magnetic field strengths can immediately be ruled out, leaving behind objects that are known to exhibit large scale variability. 
 
Although the CFHT optical imaging was not taken at the same time as \cf\ was detected in X-rays, we would still expect to detect an optical counterpart consistent with the X-ray source position if \cf\ were a star, a cataclysmic variable (CV) or an Active Galactic Nucleus (AGN). The X-ray detection and deep $g$-band and $i$-band limits indicate X-ray to optical flux ratios of F$_X$/F$_g$ $\geq$ 8200 and F$_X$/F$_i$ $\geq$ 5800 for \cf. Even if the source just outside the \xmm\ error circle were the counterpart the flux ratios would be F$_X$/F$_g$ $\sim$ 1300 and F$_X$/F$_i$ $\sim$ 340. These are far in excess of the ratios observed for AGN and stars, which typically have X-ray to optical flux ratios of F$_X$/F$_O$ $<$ 10 and $<$ 0.01, respectively \citep[e.g.][]{mai02}. It is also inconsistent with CVs which typically have F$_X$/F$_O$ $\leq$ 0.1 \citep{kuu06}. 

High levels of dust extinction could explain the lack of an optical counterpart\footnote{But not the lack of a NIR counterpart in the 2MASS images.}, however the position of \cf\ $\sim$22$^\circ$ out of the Galactic plane puts it in a region of low dust extinction with E(B-V) = 0.062 mag \citep{sch98}. The weighted average Galactic neutral hydrogen absorption in the direction of \cf\ as measured by the Leiden/Argentine/Bonn (LAB) survey of Galactic HI is 7.6 $\times$ 10$^{20}$ cm$^{-2}$ \citep{kal05}, consistent with the N$_H$ values determined through the X-ray spectral fitting (i.e. 8 -- 35 $\times$ 10$^{20}$ cm$^{-2}$, Tables \ref{bhxrb} \& \ref{specpar}) with some additional intrinsic absorption\footnote{We note that the N$_H$ of the blackbody plus power law model was higher than the other models and a factor of $\sim$5 above the Galactic value (still consistent with a moderate level of intrinsic absorption). The N$_H$ and power law photon index are degenerate in the fit, so the parameters of this model are likely to be untrustworthy.}.
If \cf\ is located within or behind M 31 there may be an additional contribution from the galaxy disk. However, the low N$_H$ values from the X-ray spectral fitting indicate that M 31 is unlikely to be a major contributor to the extinction. We thus conclude that the high flux ratio is not due to dust reddening. 

The lack of a point-like radio counterpart above 1.5 mJy is not very constraining. For example, using the black hole fundamental plane relationship between X-ray luminosity, radio luminosity, and black hole mass, we would not expect to be able to detect radio emission from an AGN unless it was closer than 1.5 Mpc \citep[assuming a black hole mass of 10$^6$ M$_\odot$; e.g.][]{koe06}. At such a distance the galaxy itself should be easily resolved in the optical/NIR images. Likewise, radio emission from the jet of a 20 M$_\odot$ black hole X-ray binary would not be detectable unless the system was closer than 20 pc, a distance at which  emission from the disk or donor star would surely be detectable in our deep optical images.  


The EPIC X-ray spectra are consistent with blackbody plus steep power law, double blackbody, disk blackbody plus power law, disk blackbody plus thermal Comptonisation, or thermal bremsstrahlung models (all with the addition of a photoelectric absorption component of low column density, consistent with the Galactic latitude of the source). The temperature of the bremsstrahlung model was 0.98 keV, much lower than the lowest temperature measured for a CV \citep[kT = 1.9 keV from thermal plasma fits to \xmm\ spectra of the quiescent dwarf nova GW Lib;][]{hil07}. In addition, line emis- sion becomes more significant relative to the un- derlying continuum spectrum for CVs with low temperature spectra. Hence, if XMM J0038+40 was a CV it would be expected that the spectrum would be consistent with a thermal plasma model with non-zero elemental abundances. The spectrum of \cf\ is not consistent with such a model, adding further weight to the argument against it being a CV.

Spectra of accreting black hole and neutron star X-ray binaries are typically modelled by a power law (representing inverse Compton emission), sometimes with the addition of a soft thermal component attributed either to emission from the surface (in the case of a neutron star) or the accretion disk \citep[e.g.][]{rem06,orl06}. Observationally it can be difficult to discriminate between black hole and neutron stars in X-ray binaries unless the compact object undergoes behaviour such as thermonuclear bursts or spin period pulsations that are definitively identified with neutron stars. However, a clear separation between neutron star and black hole X-ray binaries has been found empirically in the \emph{XMM-Newton} hardness ratios, particularly evident in the medium and hard X-ray bands (Farrell et al. 2012, in preparation). This separation was previously noted by \citet{don07} in \emph{RXTE} data and attributed to the presence of a surface in neutron stars and an event horizon in black holes. When compared with a sample of neutron star and black hole X-ray binaries drawn from the 2XMM catalogue the hardness ratios of \cf\ place it in the parameter space populated by black hole X-ray binaries. The only neutron star systems that have spectra as soft are the rare class of quiescent neutron star X-ray binaries. The spectra of these objects are dominated by thermal emission from the surface or atmosphere of the neutron star \citep[with a faint Compton tail;][and references therein]{cam04}, which is inconsistent with the spectral fitting of \cf\ (see Table \ref{specpar}).



We now consider the possibility that \cf\ is a black hole X-ray binary. The temperature derived for the disk blackbody components when fitting the EPIC spectra with the disk blackbody plus power law model (see Table \ref{bhxrb}) is significantly less than that generally seen in black hole binary spectra when the disk is inferred to be at the inner stable circular orbit \citep[ISCO; see the reviews of e.g.][]{mcc06,don07}. Such a low temperature is most consistent with black hole binaries that are not in the disk dominated spectral state but are instead transitioning to the low/hard state where the disk does not extend to the ISCO. If the disk is truncated at the lowest mass accretion rates \citep[see][]{mac03} then the temperature of this component should be lower and the spectrum should be appended by a strong, hard tail of emission \citep{mcc06}. This tail is accompanied by large amounts of variability on all timescales \citep[although this variability itself may originate in the disk: see][]{utt11}, inconsistent with the absence of variability in the EPIC light curves of \cf.

\begin{figure}
\begin{center}
\includegraphics[width=\columnwidth]{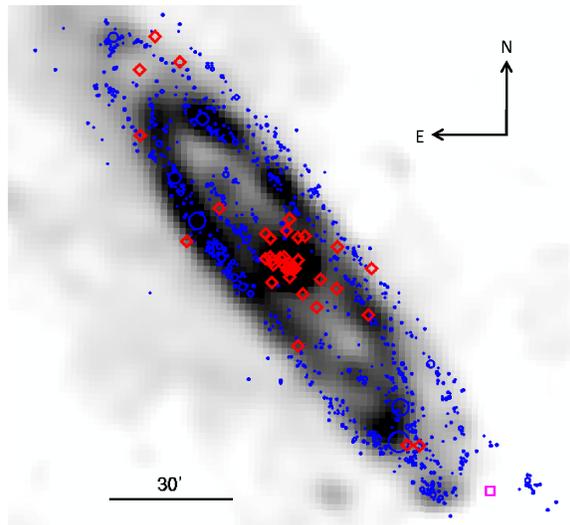}
 \caption{IRAS 60 $\mu$m image of M 31 with the distribution of star forming regions \citep[blue circles;][]{kan09} plus X-ray binaries and candidates from the XMM LP-total catalogue \citep[red diamonds;][]{sti11} over-plotted. All the X-ray binaries are located within the dusty star forming rings of M 31 except for \cf\ (purple square), which is distinctly removed from any nearby regions of star formation. }\label{xrbfig}
\end{center}
\end{figure}

While such low disk temperatures are most often seen in conjunction with variability, there are cases where it is not. One example is the black hole X-ray binary XTE J1752-223, which has been observed to have a disk temperature as low as 0.5 keV with $<$ 20\% RMS variability before beginning the transition to the low/hard state \citep{cur11}. The lack of variability and the low disk temperature does not therefore rule out \cf\ as a black hole X-ray binary. The luminosity of XTE J1752-223 while it displayed a disk temperature of $\sim$0.5 keV was $\sim$2 $\times$ 10$^{37}$ erg s$^{-1}$ in the 0.2 -- 10 keV band \citep[assuming the distance of 3.5 kpc derived by][]{sha10}. In comparison, \cf\ has a luminosity of  $\sim$2 $\times$ 10$^{38}$ erg s$^{-1}$ if it is located in M 31 (D $\sim$ 0.8 Mpc), indicating that (if it is a black hole X-ray binary) it makes the transition from the high/soft to low/hard spectral states at an Eddington fraction a factor of 10 higher than XTE J1752-223. Alternatively, if \cf\ is located within our own Galaxy, its luminosity would be $<$ 4 $\times$ 10$^{34}$ erg s$^{-1}$ (assuming a distance $<$ 15 kpc), far too low for a black hole X-ray binary in the high/soft state and therefore favouring a location within M 31. However, if \cf\ is a black hole binary in M 31 it is located in an unusual environment. All the known M 31 X-ray binaries and candidates are either located in globular clusters or within the dusty star forming rings \citep{sti11}. In contrast, \cf\ is not coincident with a globular cluster and is found well outside the star forming ring (see Figure \ref{xrbfig}), indicating that it must have been ejected from its birth location. The angular distance from \cf\ to the nearest star forming ring is $\sim$8.5\arcmin, implying a minimum distance of $\sim$2 kpc at the distance of M 31. Adopting a kick velocity of 500 km s$^{-1}$ from the supernova explosion that formed the black hole, we derive a travel time of $>$ 4 Myr, well within the project lifetime of a low mass donor star. 


We now consider the possibility that \cf\ could be a magnetar. The X-ray spectra of magnetars have been historically fitted with a two component model comprised of a steep power law plus a blackbody \citep[e.g.][]{mar01}\footnote{We caution that this is not a a physical representation, and that the power law component is invoked simply to fit the high-energy residuals that the blackbody component does not account for. While the blackbody temperature might have some physical meaning for sources with reasonable absorption, for sources with low absorption such as \cf\ the power law blows up at low-energy and the temperature of the blackbody fit is probably not meaningful.}. However, \citet{hal05} argue that this model is not physically justified and favour the fitting of a double blackbody model. They emphasize this is more physically feasible as the pulse modulation of the magnetar can be understood in terms of anisotropic heat conduction and radiative transfer in the strong magnetic field. The EPIC spectra of \cf\ are consistent with both the blackbody plus steep power law and the double blackbody models, with the spectral parameters falling well within the range spanned by SGRs and AXPs \citep[e.g.][]{mar01}. The absence of slow pulsations in the light curve of \cf\ (one of the defining characteristics of magnetars) does not weigh against the magnetar hypothesis, as pulsed fractions below the upper limit of 7$\%$ derived for \cf\ have been observed from other magnetars \citep[e.g. the AXP 4U 0142+614 has been observed with a pulsed fraction as low as $\sim$5$\%$;][]{rea07}.

Magnetars typically have very faint or no counterparts in NIR and optical wavelengths  \citep[e.g.][]{hul00,gel07}, consistent with the deep limits we have obtained for \cf. In particular, using the unabsorbed X-ray flux and the 2MASS K$_s$ band magnitude limit, the X-ray to NIR flux ratio for \cf\ is F$_X$/F$_{Ks}$ $>$ 37, which is consistent with the flux ratios of the known magnetars \citep[see e.g. Figure 5 of][]{gel07}. The X-ray variability by a factor of $\gtrsim$ 450 observed from \cf\ is of the same order as seen from other magnetars \citep[e.g. XTE J1810--197, CXOU J164710.2--455216, and AX J1845.0--0258; see e.g.][and references therein]{tam06,alb10} but slightly higher than the most extreme example \citep[AX J1845.0--0258 with a variability factor of 260 -- 430;][]{tam06}. Thus, the X-ray spectrum, transient behaviour, and high F$_X$/F$_O$ ratio are all consistent with the known sample of magnetars.

Apart from the two magnetars located in the Magellanic Clouds, all the other members of this class have been found very close to the plane of our own Galaxy. In contrast, \cf\ is coincident with the outer edges of the nearby galaxy M 31. The 0.2 -- 10 keV X-ray luminosity of \cf\ at the distance of M 31 is $\sim2 \times 10^{38}$ erg s$^{-1}$, about 3 orders of magnitude higher than typically seen from magnetars following an outburst \citep{rea11}. In addition, if \cf\ is located in M 31, the radius of the blackbody emitting radii derived from the normalization of the BBODYRAD components are 554 km and 126 km for the double blackbody model and 162 km for the blackbody plus power law model. These radii are far greater than the maximum radius of a neutron star. When taken in conjunction with the unusually high luminosity, we conclude that \cf\ is unlikely to be a magnetar located within M 31, and (if it is a magnetar) is most likely to be Galactic.



If \cf\ is $<$ 10,000 yr old with a spatial velocity $<$ 500 km s$^{-1}$, assuming it was born in a supernova explosion from a massive star in the Galactic plane \citep[e.g.][]{gae01}, then it could only have travelled $\sim$5 pc directly out of the plane. The Galactic latitude of b $\sim$22$^\circ$ therefore implies a line-of-sight distance of $\sim$13 pc, making \cf\ an improbably nearby magnetar. Furthermore, the derived luminosity at such a distance is only $\sim5 \times 10^{27}$ erg s$^{-1}$, significantly lower than typically observed from magnetars. If the luminosity of \cf\ is between 10$^{32}$ -- 10$^{35}$ erg s$^{-1}$ like the other magnetars, this implies a distance of between $\sim$0.7 -- 21 kpc from Earth and therefore a distance of $\sim$0.3 -- 8 kpc from the Galactic plane. This then implies either an implausibly high spatial velocity $>$ 10,000 km s$^{-1}$ (for an age $<$ 10,000 yr) or an age $> 3 \times 10^6$ yr (for a spatial velocity $<$ 500 km s$^{-1}$). 

Such an age is inconsistent with that derived for magnetars. However, SGR-like behaviour has been observed from a source (SGR 0418+5729) located at a moderate Galactic latitude (b = +5.1$^\circ$). Pulsations with a period of $\sim$9.1 s have been detected from SGR 0418+5729 but with a very small period derivative of \.{P} $< 6 \times 10^{-15}$ s s$^{-1}$, implying a surface dipolar magnetic field strength of $< 7.5 \times 10^{12}$ G and a characteristic age of $> 24 \times 10^6$ yr \citep{rea10}. Thus, if the progenitor of \cf\ was a massive star that went supernova within the Galactic plane, it is possible that it is an old magnetar where the magnetic field strength has decayed to $\sim10^{12}$ G.

Alternatively, the high Galactic latitude of \cf\ could be explained by the progenitor being a massive runaway star that originated in a binary system that became unbound, and then travelled outside the plane of the Galaxy prior to going supernova. Runaway O and B-type stars have been observed in our Galaxy with space velocities up to $\sim$100 km s$^{-1}$ \citep{hoo01}. Assuming a similar space velocity, the progenitor of \cf\ would have taken $\sim3 \times 10^6 - 8 \times 10^7$ yr to travel $\sim$0.3 -- 8 kpc out of the Galactic plane (assuming a velocity vector perpendicular to the plane). These timescales are perfectly consistent with the nuclear burning lifetimes of massive stars with masses between $\sim$8 -- 30 M$_\odot$ with lifetimes between $\sim7 - 40 \times 10^6$ yr \citep{mae89}. 

In order to discriminate between a black hole X-ray binary and a magnetar, we need to either detect spectral state transitions (that are the signature of black hole binaries) or pulsations (that are a signature of magnetars). If the pulsations are detected, the measurement of the spin derivative would then allow us to derive both the surface dipole magnetic field strength and the characteristic age of the neutron star, discriminating between the two possibilities we describe above for the nature of the magnetar. Further deep X-ray observations triggered based on regular monitoring of \cf\ are thus required.



\acknowledgments

We thank the anonymous referee for their comments which improved the paper, and Jeff McClintock for useful discussions. JRC acknowledges the support of a vacation scholarship from the School of Physics at The University of Sydney. SAF is the recipient of an Australian Research Council Post Doctoral Fellowship, funded by grant DP110102889. Parts of this research were conducted by the Australian Research Council Centre of Excellence for All-sky Astrophysics (CAASTRO), through project number CE110001020. GFL thanks the Australian Research Council for support through his Future Fellowship (FT100100268) and Discovery Project (DP110100678). MJM thanks STFC for their support in the form of a postdoctoral position funded by a standard grant. We thank Neil Gehrels and the $\emph{Swift}$ team for the follow-up Target of Opportunity observation. Based on observations from XMM-Newton, an ESA science mission with instruments and contributions directly funded by ESA Member States and NASA. This work made use of the 2XMM Serendipitous Source Catalogue, constructed by the XMM-Newton Survey Science Centre on behalf of ESA. We also thank the  Pan-Andromeda Archaeological Survey (PAndAS) collaboration for kindly providing the CFHT data. This publication makes use of data products from the Two Micron All Sky Survey, which is a joint project of the University of Massachusetts and the Infrared Processing and Analysis Center/California Institute of Technology, funded by the National Aeronautics and Space Administration and the National Science Foundation. The Centre for All-sky Astrophysics is an Australian Research Council Centre of Excellence, funded by grant CE11E0090. 



{\it Facilities:} \facility{XMM-Newton (EPIC)}.

\clearpage

\appendix

\section{Details of X-ray Observations}

Table \ref{obslog} lists the details of the X-ray observations of \cf\ used for our analysis. The flux limits quoted are the observed (i.e. absorbed) 0.2 -- 10 keV 3$\sigma$ limits in units of erg cm$^{-2}$ s$^{-1}$, assuming the absorbed double black body model parameters given in Table \ref{specpar}. For the 2008 \xmm\ observation in which \cf\ was detected, we give the absorbed flux in bold derived from the double blackbody model. For the two remaining \xmm\ observations, the upper limit was only calculated for the pn image. The images for the \emph{Swift} data taken in 2011 were summed together to increase signal-to-noise, so the upper limits are for the combined data. While the upper limits are quoted for each individual \emph{ROSAT} observation, the limit from the summed observations is 3.9 $\times$ 10$^{-15}$ erg cm$^{-2}$ s$^{-1}$.

\begin{deluxetable}{cccccc}
\tablewidth{0pt}
\tablecaption{Log of X-ray Observations.}\label{obslog}
\tablehead{
\colhead{Telescope} & \colhead{ObsID}           & \colhead{Date}      &
\colhead{Instrument}            &
\colhead{T$_{exp}$\tablenotemark{a}} & \colhead{Flux Upper Limit}        
}
\startdata
\emph{ROSAT}	&	600079N00	&	1991-07-14	&	PSPC	&	41.7	&	1.7	$\times$ 10$^{-14}$	\\
\emph{ROSAT}	&	600064N00	&	1991-07-15	&	PSPC	&	48.8	&	2.8	$\times$ 10$^{-14}$	\\
\emph{ROSAT}	&	141836N00	&	1992-01-12	&	PSPC	&	1.7	&	6.7	$\times$ 10$^{-14}$	\\
\emph{ROSAT}	&	141839N00	&	1992-01-13	&	PSPC	&	1.8	&	3.6	$\times$ 10$^{-14}$	\\
\emph{ROSAT}	&	141837N00	&	1992-01-13	&	PSPC	&	2.1	&	1.2	$\times$ 10$^{-13}$	\\
\emph{ROSAT}	&	600317N00	&	1992-07-20	&	PSPC	&	2.6	&	5.4	$\times$ 10$^{-14}$	\\
\emph{ROSAT}	&	600301N00	&	1992-07-22	&	PSPC	&	2.7	&	3.9	$\times$ 10$^{-14}$	\\
\emph{ROSAT}	&	600341N00	&	1992-07-23	&	PSPC	&	2.7	&	9.3	$\times$ 10$^{-14}$	\\
\emph{ROSAT}	&	600360N00	&	1992-07-23	&	PSPC	&	2.9	&	1.2	$\times$ 10$^{-13}$	\\
\emph{ROSAT}	&	600339N00	&	1992-07-25	&	PSPC	&	2.8	&	7.9	$\times$ 10$^{-14}$	\\
\emph{ROSAT}	&	600296N00	&	1992-07-25	&	PSPC	&	2.5	&	8.7	$\times$ 10$^{-14}$	\\
\emph{ROSAT}	&	600297N00	&	1992-07-25	&	PSPC	&	2.5	&	1.2	$\times$ 10$^{-13}$	\\
\emph{ROSAT}	&	600358N00	&	1992-07-26	&	PSPC	&	2.9	&	4.1	$\times$ 10$^{-14}$	\\
\emph{ROSAT}	&	600303N00	&	1992-07-27	&	PSPC	&	2.4	&	9.2	$\times$ 10$^{-14}$	\\
\emph{ROSAT}	&	600321N00	&	1992-07-29	&	PSPC	&	2.8	&	6.9	$\times$ 10$^{-14}$	\\
\emph{ROSAT}	&	600336N00	&	1992-07-29	&	PSPC	&	2.8	&	8.8	$\times$ 10$^{-14}$	\\
\emph{ROSAT}	&	600356N00	&	1992-08-03	&	PSPC	&	2.0	&	2.5	$\times$ 10$^{-14}$	\\
\emph{ROSAT}	&	600338N00	&	1992-08-03	&	PSPC	&	2.2	&	5.6	$\times$ 10$^{-14}$	\\
\emph{ROSAT}	&	600361N00	&	1992-08-05	&	PSPC	&	2.7	&	4.8	$\times$ 10$^{-14}$	\\
\emph{ROSAT}	&	600342N00	&	1992-08-05	&	PSPC	&	3.4	&	6.8	$\times$ 10$^{-14}$	\\
\emph{ROSAT}	&	600300N00	&	1992-08-05	&	PSPC	&	2.5	&	8.6	$\times$ 10$^{-14}$	\\
\emph{ROSAT}	&	600323N00	&	1992-08-05	&	PSPC	&	2.6	&	1.5	$\times$ 10$^{-13}$	\\
\emph{ROSAT}	&	600298N00	&	1992-08-05	&	PSPC	&	2.6	&	2.2	$\times$ 10$^{-13}$	\\
\emph{ROSAT}	&	600362N00	&	1992-08-06	&	PSPC	&	2.4	&	5.7	$\times$ 10$^{-14}$	\\
\emph{ROSAT}	&	600319N00	&	1992-08-06	&	PSPC	&	1.8	&	5.8	$\times$ 10$^{-14}$	\\
\emph{ROSAT}	&	600340N00	&	1992-08-06	&	PSPC	&	2.6	&	6.2	$\times$ 10$^{-14}$	\\
\emph{ROSAT}	&	600322N00	&	1992-08-06	&	PSPC	&	2.6	&	1.0	$\times$ 10$^{-13}$	\\
\emph{ROSAT}	&	600318N00	&	1992-08-06	&	PSPC	&	1.4	&	1.2	$\times$ 10$^{-13}$	\\
\emph{ROSAT}	&	600359N00	&	1992-08-06	&	PSPC	&	2.6	&	1.7	$\times$ 10$^{-13}$	\\
\emph{ROSAT}	&	600318a01	&	1992-12-21	&	PSPC	&	1.3	&	1.2	$\times$ 10$^{-13}$	\\
\emph{ROSAT}	&	600316N00	&	1992-12-21	&	PSPC	&	2.7	&	1.4	$\times$ 10$^{-13}$	\\
\emph{ROSAT}	&	600357N00	&	1992-12-30	&	PSPC	&	2.7	&	4.1	$\times$ 10$^{-14}$	\\
\emph{ROSAT}	&	600302N00	&	1992-12-30	&	PSPC	&	2.6	&	1.3	$\times$ 10$^{-13}$	\\
\emph{ROSAT}	&	600337N00	&	1992-12-30	&	PSPC	&	2.1	&	1.5	$\times$ 10$^{-13}$	\\
\emph{ROSAT}	&	600299a01	&	1993-01-01	&	PSPC	&	1.4	&	6.2	$\times$ 10$^{-14}$	\\
\emph{ROSAT}	&	600244N00	&	1993-01-02	&	PSPC	&	35.9	&	1.8	$\times$ 10$^{-14}$	\\
\emph{ROSAT}	&	600343N00	&	1993-01-04	&	PSPC	&	2.7	&	7.5	$\times$ 10$^{-14}$	\\
\emph{ROSAT}	&	600320a01	&	1993-07-01	&	PSPC	&	2.7	&	6.3	$\times$ 10$^{-14}$	\\
\emph{Chandra}	&	2046	&	2000-11-05	&	ACIS-S	&	14.8	&	1.6	$\times$ 10$^{-14}$	\\
\emph{Chandra}	&	2047	&	2001-03-06	&	ACIS-S	&	14.8	&	2.1	$\times$ 10$^{-14}$	\\
\emph{Chandra}	&	2048	&	2001-07-03	&	ACIS-S	&	14.0	&	1.4	$\times$ 10$^{-14}$	\\
\emph{XMM-Newton}	&	0402560101	&	2006-06-28	&	pn	&	44.1	&	4.2	$\times$ 10$^{-15}$	\\
\emph{Swift}	&	00035337001	&	2007-06-01	&	XRT	&	9.2	&	9.0	$\times$ 10$^{-15}$	\\
\emph{XMM-Newton}	&	0505760101	&	2007-07-24	&	pn	&	57.0	&	3.2	$\times$ 10$^{-15}$	\\
\emph{XMM-Newton}	&	0511380101	&	2008-01-02	&	pn	&	44.1	&	\textbf{1.5	$\times$ 10$^{-12}$}	\\
\emph{Swift}	&	00031919001	&	2011-02-04	&	XRT	&	2.9	&	3.3	$\times$ 10$^{-14}$	\\
\emph{Swift}	&	00031919002	&	2011-02-07	&	XRT	&	1.4	&	3.3	$\times$ 10$^{-14}$	\\
\emph{Swift}	&	00031919003	&	2011-02-11	&	XRT	&	0.9	&	3.3	$\times$ 10$^{-14}$	\\
\emph{Swift}	&	00031919004	&	2011-02-15	&	XRT	&	0.7	&	3.3	$\times$ 10$^{-14}$	\\
           

\enddata
\tablenotetext{a}{Exposure time in kiloseconds.}

\end{deluxetable}


\begin{thebibliography}{}

\bibitem[Albano et al.(2010)]{alb10} Albano, A., Turolla, R., Israel, G.~L., Zane, S., Nobili, L., \& Stella, L.\ 2010, \apj, 722, 788 
\bibitem[Arnaud(1996)]{arn96} Arnaud, K. A. 1996, Astronomical Data Analysis Software and Systems V, 101, 17 
\bibitem[Bradley et al.(2007)]{bra07} Bradley, C.~K., Hynes, R.~I., Kong, A.~K.~H., Haswell, C.~A., Casares, J., \& Gallo, E.\ 2007, \apj, 667, 427 
\bibitem[Bystedt et  al.(1984)]{bys84} Bystedt, J.~E.~V., Brinks, E., de Bruyn, A.~G., Israel, F.~P., Schwering, P.~B.~W., Shane, W.~W., \& Walterbos, R.~A.~M.\ 1984, \aaps, 56, 245 
\bibitem[Campana \& Stella(2004)]{cam04} Campana, S., \& Stella, L.\ 2004, NuPhS, 132, 427 
\bibitem[Condon et al.(1998)]{con98} Condon, J.~J., Cotton, W.~D., Greisen, E.~W., Yin, Q.~F., Perley, R.~A., Taylor, G.~B., 
\& Broderick, J.~J.\ 1998, \aj, 115, 1693 
\bibitem[Curran et al.(2011)]{cur11} Curran, P.~A., Maccarone, T.~J., Casella, P., et al.\ 2011, \mnras, 410, 541 
\bibitem[Duncan \& Thompson(1992)]{dun92} Duncan, R.~C. \& Thompson, C. 1992, ApJ, 392, L9
\bibitem[Done et al.(2007)]{don07} Done, C., Gierli{\'n}ski, M., \& Kubota, A.\ 2007, \aapr, 15, 1 
\bibitem[Edelson et al.(2002)]{ede02} Edelson, R., Turner, T.~J., Pounds, K., Vaughan, S., Markowitz, A., Marshall, H., Dobbie, P., \& Warwick, R.\ 2002, \apj, 568, 610 
\bibitem[Evans et al.(2009)]{eva09} Evans, P.~A., et al.\ 
2009, \mnras, 397, 1177
\bibitem[Farrell et al.(2010)]{far10} Farrell, S.~A., Gosling, A.~J., Webb, N.~A., Barret, D., Rosen, S.~R., Sakano, M., \& Pancrazi, B.\ 2010, \aap, 523, A50 
\bibitem[Frederiks et al.(2007)]{fre07} Frederiks, D.~D., Palshin, V.~D., Aptekar, R.~L., Golenetskii, S.~V., Cline, T.~L., \& Mazets, E.~P. 2007, AstL, 33, 19 
\bibitem[Gaensler et al.(2001)]{gae01} Gaensler, B.~M., Slane, P.~O., Gotthelf, 
E.~V., \& Vasisht, G. 2001, ApJ, 559, 963 
\bibitem[Gavriil, Kaspi, \& Woods(2002)]{gav02} Gavriil, F.~P., Kaspi, V.~M., \& Woods, P.~M. 2002, Nature, 419, 142 
\bibitem[Gelfand et al.(2004)]{gel04} Gelfand, J.~D., Lazio, T.~J.~W., \& Gaensler, B.~M.\ 2004, \apjs, 155, 89 
\bibitem[Gelfand \& Gaensler(2007)]{gel07} Gelfand, J.~D., \& Gaensler, B.~M.\ 2007, \apj, 667, 1111 
\bibitem[Halpern \& Gotthelf(2005)]{hal05} Halpern, J.~P. \& Gotthelf, E.~V. 2005, ApJ, 618, 874 
\bibitem[Hilton et al.(2007)]{hil07} Hilton, E.~J., Szkody, P., Mukadam, A., Mukai, K., Hellier, C., van Zyl, L., 
\& Homer, L.\ 2007, \aj, 134, 1503 
\bibitem[Hoogerwerf et al.(2001)]{hoo01} Hoogerwerf, R., de Bruijne, J.~H.~J., \& de Zeeuw, P.~T.\ 2001, \aap, 365, 49 
\bibitem[Hulleman et al.(2000)]{hul00} Hulleman, F., van Kerkwijk, M.~H., \& Kulkarni, S.~R.\ 2000, \nat, 408, 689 
\bibitem[Hynes et al.(2004)]{hyn04} Hynes, R.~I., et al.\ 2004, \apjl, 611, L125 
\bibitem[Ibata et al.(2007)]{iba07} Ibata, R., Martin, N.~F., Irwin, M., Chapman, S., Ferguson, A.~M.~N., Lewis, G.~F., \& McConnachie, A.~W.\ 2007, \apj, 671, 1591 
\bibitem[Kaastra (1992)]{kaa92} ]Kaastra, J.S. 1992, An X-Ray Spectral Code for Optically Thin Plasmas (Internal SRON-Leiden Report, updated version 2.0)
\bibitem[Kalberla et al.(2005)]{kal05} Kalberla, P.~M.~W., Burton, W.~B., Hartmann, D., Arnal, E.~M., Bajaja, E., Morras, R., P\"{o}ppel, W.~G.~L.\ 2005, \aap, 440, 775 
\bibitem[Kang et al.(2009)]{kan09} Kang, Y., Bianchi, L., \& Rey, S.-C.\ 2009, \apj, 703, 614
\bibitem[Kaspi et al.(2003)]{kas03} Kaspi, V.~M., Gavriil, F.~P., Woods, P.~M., Jensen, J.~B., Roberts, M.~S.~E., \& 
Chakrabarty, D. 2003, ApJ, 588, L93 
\bibitem[K{\"o}rding et al.(2006)]{koe06} K{\"o}rding, E., Falcke, H., \& Corbel, S.\ 2006, \aap, 456, 439 
\bibitem[Kouveliotou et al.(1998)]{kou98} Kouveliotou, C., et al. 1998, Nature, 393, 235 
\bibitem[Kuulkers et al.(2006)]{kuu06} Kuulkers, E., Norton, A., Schwope, A., \& Warner, B.\ 2006, Compact stellar X-ray sources, 421 
\bibitem[Leahy et al.(1983)]{lea83} Leahy, D.~A., Elsner, R.~F., \& Weisskopf, M.~C.\ 1983, \apj, 272, 256 
\bibitem[Liedahl et al.(1995)]{lie95} Liedahl, D.~A., Osterheld, A.~L., \& Goldstein, W.~H.\ 1995, \apjl, 438, L115 
\bibitem[Maccarone(2003)]{mac03} Maccarone, T.~J.\ 2003, \aap, 409, 697 
\bibitem[Maeder \& Meynet(1989)]{mae89} Maeder, A., \& Meynet, G.\ 1989, \aap, 210, 155 
\bibitem[Mainieri et al.(2002)]{mai02} Mainieri, V., Bergeron, J., Hasinger, G., Lehmann, I., Rosati, P., Schmidt, M., Szokoly, G., \& Della Ceca, R.\ 2002, \aap, 393, 425 
\bibitem[Marsden \& White(2001)]{mar01} Marsden, D., \& White, N.~E.\ 2001, \apjl, 551, L155
\bibitem[Mazets et al.(1979)]{maz79} Mazets, E.~P., Golentskii, S.~V., Ilinskii, V.~N., Aptekar, R.~L., \& Guryan, Iu.~A. 1979, \nat, 282, 587
\bibitem[Mazets et al.(2008)]{maz08} Mazets, E.~P., et al. 2008, ApJ, 680, 545 
\bibitem[McClintock \& Remillard(2006)]{mcc06} McClintock, J.~E., \& Remillard, R.~A.\ 2006, Compact stellar X-ray sources, 157 
\bibitem[McConnachie et al.(2009)]{mcc09} McConnachie, A.~W., et al.\ 2009, \nat, 461, 66 
\bibitem[McGarry et al.(2005)]{mcg05} McGarry, M.~B., Gaensler, B.~M., Ransom, S.~M., Kaspi, V.~M., \& Veljkovik, S.\ 2005, \apjl, 627, L137 
\bibitem[Mewe et al.(1985)]{mew85} Mewe, R., Gronenschild, E.~H.~B.~M., \& van den Oord, G.~H.~J.\ 1985, \aaps, 62, 197 
\bibitem[Mewe et al.(1986)]{mew86} Mewe, R., Lemen, J.~R., \& van den Oord, G.~H.~J.\ 1986, \aaps, 65, 511 
\bibitem[Narayan et al.(1997)]{nar97} Narayan, R., Kato, S., \& Honma, F.\ 1997, \apj, 476, 49 
\bibitem[Ofek et al.(2008)]{ofe08} Ofek, E.~O., et al. 2008, ApJ, 681, 1464 
\bibitem[Orlandini(2006)]{orl06} Orlandini, M.\ 2006, AdSpR, 38, 2742 
\bibitem[Rea et al.(2007)]{rea07} Rea, N., et al.\ 2007, \mnras, 381, 293 
\bibitem[Rea et al.(2010)]{rea10} Rea, N., et al.\ 2010, Science, 330, 944 
\bibitem[Rea \& Esposito(2011)]{rea11} Rea, N., \& Esposito, P.\ 2011, High-Energy Emission from Pulsars and their Systems, 247 
\bibitem[Remillard \& McClintock(2006)]{rem06} Remillard, R.~A., \& McClintock, J.~E.\ 2006, \araa, 44, 49 
\bibitem[Rengelink et al.(1997)]{ren97} Rengelink, R.~B., Tang, Y., de Bruyn, A.~G., Miley, G.~K., Bremer, M.~N., Roettgering, H.~J.~A., \& Bremer, M.~A.~R.\ 1997, \aaps, 124, 259 
\& Zavlin, V.~E.\ 2000, \apj, 529, 985 
\bibitem[Schlegel et al.(1998)]{sch98} Schlegel, D.~J., Finkbeiner, D.~P., \& Davis, M.\ 1998, \apj, 500, 525 
\bibitem[Shaposhnikov et al.(2010)]{sha10} Shaposhnikov, N., Markwardt, C., Swank, J., \& Krimm, H.\ 2010, \apj, 723, 1817 
\bibitem[Stiele et al.(2011)]{sti11} Stiele, H., Pietsch, W., Haberl, F., et al.\ 2011, \aap, 534, A55 
\bibitem[Tam et al.(2006)]{tam06} Tam, C.~R., Kaspi, V.~M., Gaensler, B.~M., \& Gotthelf, E.~V.\ 2006, \apj, 652, 548 
\bibitem[Uttley et al.(2011)]{utt11} Uttley, P., Wilkinson, T., Cassatella, P., Wilms, J., Pottschmidt, K., Hanke, M., B\"{o}ck, M.\ 2011, \mnras, 414, L60 
\bibitem[van der Klis(1989)]{van89} van der Klis, M. 1989, in Timing neutron stars, ed. H. \"{O}gelman \& E. P. J. van den Heuvel (New York, USA: Kluwer Academic/Plenum Publishers), 27
\bibitem[Vaughan et al.(2003)]{vau03} Vaughan, S., Edelson, R., Warwick, R.~S., \& Uttley, P.\ 2003, \mnras, 345, 1271 
\bibitem[Vaughan(2010)]{vau10} Vaughan, S.\ 2010, \mnras, 402, 307 
\bibitem[Walterbos et al.(1985)]{wal85} Walterbos, R.~A.~M., Brinks, E., \& Shane, W.~W.\ 1985, \aaps, 61, 451 
\bibitem[Watson et al.(2009)]{wat09} Watson, M.~G., et al. 2009, A\&A, 493, 339
\bibitem[Wilms, Allen, \& McCray(2000)]{wil00} Wilms, J., Allen, A. \& McCray, R. 2000, ApJ, 542, 914 
\bibitem[Woods \& Thompson(2006)]{woo06} Woods, P.~M., \& Thompson, C.\ 2006, in Compact stellar X-ray sources, ed. W. Lewin \& M. van der Klis (Cambridge, UK: Cambridge University Press), 547 



\end{thebibliography}
\end{document}